\documentclass[prd,aps,showpacs,notitlepage,superscriptaddress]{revtex4-1}
\usepackage{amsmath}
\usepackage{epsfig}
\newcommand{\ben}{\begin{eqnarray}}
\newcommand{\een}{\end{eqnarray}}
\newcommand{\nnu}{\nonumber\\}
\newcommand{\bef}{\begin{figure}[htb]\centering}
\newcommand{\eef}{\end{figure}}
\newcommand{\hx}{\hat{x}}
\newcommand{\hz}{\hat{z}}
\newcommand{\pht}{P_{h\perp}}

\begin{document}
\title{Transverse momentum-weighted Sivers asymmetry 
\\ in semi-inclusive deep inelastic scattering at next-to-leading order}

\date{\today}

\author{Zhong-Bo Kang}
\email{zkang@lanl.gov}
\affiliation{Theoretical Division,
                   Los Alamos National Laboratory,
                   Los Alamos, NM 87545, USA}

\author{Ivan Vitev}
\email{ivitev@lanl.gov}
\affiliation{Theoretical Division,
                   Los Alamos National Laboratory,
                   Los Alamos, NM 87545, USA}

\author{Hongxi Xing}
\email{xinghx@iopp.ccnu.edu.cn}
\affiliation{Interdisciplinary Center for Theoretical Study and Department of Modern Physics,
                   University of Science and Technology of China,
                   Hefei 230026, China}
\affiliation{Institute of Particle Physics,
                   Central China Normal University,
                   Wuhan 430079, China}

\begin{abstract}
We study the next-to-leading order perturbative QCD corrections to the transverse momentum-weighted 
Sivers asymmetry in semi-inclusive hadron production in lepton-proton deep inelastic scattering.
The corresponding differential cross section is evaluated as a convolution of a twist-three 
quark-gluon correlation function, often referred to as Qiu-Sterman function, the usual 
unpolarized fragmentation function, and a hard coefficient function. By studying 
the collinear divergence structure, we identify the evolution kernel for the 
Qiu-Sterman function. The hard coefficient function, which is finite and free 
of any divergence, is evaluated at one-loop order.
\end{abstract}

\pacs{12.38.Bx, 12.39.St, 13.85.Hd, 13.88.+e}

\maketitle

\section{Introduction}
In recent years, transverse spin physics has attracted tremendous attention from both the experimental and 
theoretical communities. As transverse spin can correlate with the transverse momentum of the partons 
inside a polarized proton, transverse spin observables, such as single transverse spin asymmetries (SSAs), 
are sensitive probes of parton's transverse motion and a path to three-dimensional proton tomography~\cite{Boer:2011fh}. 
Significant theoretical progress has been made in studying the single transverse spin asymmetries in the past 
several years. Two QCD mechanisms for generating SSAs have been proposed and applied extensively 
to  phenomenology: the transverse momentum dependent (TMD) factorization 
approach~\cite{TMD-fac,Brodsky,MulTanBoe,Boer:1997nt, Anselmino:2009st} and the collinear twist-3 
factorization approach~\cite{Efremov,qiu,koike,Koike:2011mb,Liang:2012rb}. They were shown to be closely 
related  and, thus, provide a unified picture for the SSAs~\cite{unify}. 

One of the most studied asymmetries is the so-called Sivers effect~\cite{Siv90}. At the partonic level, 
it corresponds to an azimuthal correlation $\sim S_\perp\cdot (P\times k_\perp)$, with $S_\perp$ and $P$  being
the spin and momentum vector of the polarized proton and $k_\perp$ being the transverse momentum of the 
parton. Such correlation is encoded in the so-called Sivers function, if one uses the TMD factorization 
formalism; or the twist-3 quark-gluon correlation function, often referred to as Qiu-Sterman function, 
within the collinear twist-3 factorization formalism. The evolution equations for either  the Sivers 
function~\cite{Kang:2011mr,Aybat:2011zv,Anselmino:2012aa} or the Qiu-Sterman 
function~\cite{Kang:2008ey,Zhou:2008mz,Vogelsang:2009pj,Braun:2009mi,Kang:2010xv,Kang:2012em,Schafer:2012ra,Ma:2012xn} 
have been derived recently, which enhances the accuracy of the  phenomenological applications. 

A natural step forward, as a follow-up to the derivation of evolution equations, will be the computation 
of the next-to-leading order (NLO) corrections to the transverse spin-dependent cross sections. Although 
tremendous progress has been made in the evaluation of NLO perturbative QCD (pQCD) corrections to 
the spin-averaged cross sections, similar efforts on the transverse spin-dependent cross sections 
are still rather limited. This is largely due to the complexity of such type of calculations. So far, 
the only NLO correction in this direction is performed for Drell-Yan production \cite{Vogelsang:2009pj}. 
NLO corrections to other processes will provide process-dependent corrections to the hard-part 
coefficient functions, and can also be used to extract the universal behavior of the evolution kernel 
for the relevant spin-dependent parton distributions and/or fragmentation functions. A NLO calculation 
for a particular physical process, thus, provides a direct test of QCD factorization for the associated observables. 

In this paper  we  follow Ref.~\cite{Vogelsang:2009pj} on the Drell-Yan production and perform a 
NLO calculation for the $\pht$-weighted Sivers asymmetry in semi-inclusive deep inelastic scattering (SIDIS). 
Here, $\pht$ is the transverse momentum of the  final-state hadron. 
Since the transverse momentum is being integrated out, our result is presented within the collinear 
factorization formalism in terms of twist-3 Qiu-Sterman function, NLO hard-part coefficient function, 
and the usual unpolarized fragmentation function, as we  demonstrate in detail below. The rest of 
our paper is organized as follows. In Sec. II  we introduce the notation for the semi-inclusive
hadron production in deep inelastic scattering and present the $\pht$-weighted Sivers asymmetry 
at leading order. In Sec. III we present the NLO pQCD corrections. We first give the result for 
virtual corrections and then study the real corrections. We then combine the real and virtual  parts
to obtain the final cross section. We show that all the soft divergences cancel out between real 
and virtual corrections. The remaining collinear divergence is absorbed in the redefinition of the 
unpolarized fragmentation function, and the twist-3 Qiu-Sterman function. This provides an alternative 
way to derive the evolution equation for the Qiu-Sterman function. We conclude our paper in Sec. IV.

\section{Transverse Momentum-Weighted Sivers asymmetry at leading order}
We start this section by specifying our notation and the kinematics of SIDIS. We consider the scattering 
of an unpolarized lepton $e$ with momentum $\ell$, on a transversely polarized proton $p$ with momentum 
$P$ and transverse spin vector $S_\perp$,
\ben
e(\ell)+p(P, S_\perp)\to e(\ell')+h(P_h)+X,
\een
where $h$ represents the observed final-state meson with momentum $P_h$. In the approximation of 
one-photon exchange, we define the virtual photon momentum $q=\ell-\ell'$ and its invariant mass $Q^2=-q^2$. 
The usual SIDIS variables are defined as follows:
\ben
S=(P+\ell)^2, \qquad
x_B=\frac{Q^2}{2P\cdot q}, \qquad
y=\frac{P\cdot q}{P\cdot \ell} = \frac{Q^2}{x_B S}, \qquad
z_h=\frac{P\cdot P_h}{P\cdot q}.
\een
The differential cross section that includes  the so-called Sivers effect, the $\sin(\phi_h-\phi_s)$ module,  
can be written as the following form~\cite{Kang:2012xf}
\ben
\frac{d\sigma_{\rm Sivers}}{dx_B dy dz_h d^2\pht} = \sigma_0 \left [ F_{UU} + \sin(\phi_h-\phi_s) 
F_{UT}^{\sin(\phi_h-\phi_s)}\right],
\een
where $\sigma_0=\frac{2\pi\alpha_{\rm em}^2}{Q^2} \frac{1+(1-y)^2}{y}$, 
$F_{UU}$ and $F_{UT}^{\sin(\phi_h-\phi_s)}$ are the spin-averaged and transverse spin-dependent structure functions, 
respectively.
$\phi_h$ and $\phi_s$ are the azimuthal angles for the final-state hadron momentum $P_h$ and spin 
vector $S_\perp$. In this paper, we define all our angles in the so-called {\it hadron frame}~\cite{sidis}.
The spin-averaged differential cross section $d\sigma/{dx_B dy dz_h d^2\pht} \equiv \sigma_0 F_{UU}$, and 
the $\pht$-integrated 
spin-averaged cross section is defined as
\ben
\frac{d\sigma}{dx_B dy dz_h}
\equiv
\int d^2\pht \frac{d \sigma}{dx_B dy dz_hd^2\pht}.
\een
At the same time, the transverse spin-dependent differential cross section 
$d\Delta\sigma(S_\perp)/{dx_B dy dz_h d^2\pht}  \equiv \sigma_0  \sin(\phi_h-\phi_s) F_{UT}^{\sin(\phi_h-\phi_s)}$, 
and the transverse momentum-weighted transverse spin-dependent cross section is given by~\cite{Boer:1997nt}
\ben
\frac{d\langle \pht \Delta\sigma(S_\perp)\rangle}{dx_B dy dz_h}
\equiv
\int d^2\pht \epsilon^{\alpha\beta} S_{\perp}^\alpha \pht^{\beta} \frac{d \Delta\sigma(S_\perp)}{dx_B dy dz_hd^2\pht},
\een
where $\epsilon^{\alpha\beta}$ is a two-dimensional anti-symmetric tensor with $\epsilon^{12}=1$, 
and $\epsilon^{\alpha\beta} S_{\perp}^\alpha \pht^{\beta} = \pht \sin(\phi_h-\phi_s)$. Thus,  the 
$\pht$-weighted Sivers asymmetry is given by
\ben
A_{UT}^{\pht\sin(\phi_h-\phi_s)}(x_B, y, z_h) \equiv \left.\frac{d\langle \pht \Delta\sigma(S_\perp)\rangle}{dx_B dy dz_h}
\right/ \frac{d\sigma}{dx_B dy dz_h}.
\label{sivers}
\een
There are multiple studies on the higher order pQCD corrections to the spin-averaged 
cross sections, see, for example, Ref.~\cite{Altarelli:1979ub}.  As a warm-up excercise, we also 
calculate this  cross section to NLO order, and our findings are consistent with Ref.~\cite{Altarelli:1979ub}. 
Since this result is well-known, we only give the final expression here (in the $\overline{\rm MS}$ scheme)
\ben
\frac{d\sigma}{dx_B dy dz_h}
&=&\sigma_0\sum_q e_q^2
\int \frac{dx}{x} \frac{dz}{z} q(x, \mu^2) D_{q\to h}(z, \mu^2) \delta(1-\hat x)\delta(1-\hat z)
+\sigma_0 \frac{\alpha_s}{2\pi} \sum_q e_q^2 \int \frac{dx}{x} \frac{dz}{z} 
q(x, \mu^2) D_{q\to h}(z, \mu^2)
\nnu
&&
\times
\Bigg\{
\ln\left(\frac{Q^2}{\mu^2}\right)
\left[P_{qq}(\hx)\delta(1-\hz)+ P_{qq}(\hz)\delta(1-\hx) \right]
+C_F\left[\frac{1+(1-\hx-\hz)^2}{(1-\hx)_+(1-\hz)_+} - 8\delta(1-\hx)\delta(1-\hz)\right]
\nnu
&&
+\delta(1-\hz) C_F\left[(1+\hx^2) \left(\frac{\ln(1-\hx)}{1-\hx}\right)_+ 
-\frac{1+\hx^2}{1-\hx}\ln\hx+(1-\hx)
\right]
\nnu
&&
+\delta(1-\hx) C_F\left[(1+\hz^2) \left(\frac{\ln(1-\hz)}{1-\hz}\right)_+ 
+\frac{1+\hz^2}{1-\hz}\ln\hz+(1-\hz)
\right]
\Bigg\},
\label{spin-avg}
\een
where   $\hat x= x_B/x$ and $\hat z=z_h/z$  and $P_{qq}(x)$ is the usual spin-averaged quark-to-quark splitting kernel
\ben
P_{qq}(x) = C_F \left[\frac{1+x^2}{(1-x)_+}+\frac{3}{2} \delta(1-x) \right].
\label{Pqq}
\een

Let us now concentrate on the NLO correction to the transverse spin-dependent differential cross section. 
This will allow us  to study the $\pht$-weighted Sivers asymmetry, as in Eq.~\eqref{sivers}, 
at  next-to-leading order. We will start with the leading order calculation for the transverse 
spin-dependent cross section. At this order, the final-state hadron is produced through the hadronization 
of the quark, which comes from the virtual-photon quark scattering. In order to obtain a non-vanishing 
$\pht$-weighted transverse spin-dependent cross section $\langle \pht \Delta\sigma(S_\perp)\rangle$, we have 
to include the final-state multiple interactions, as shown in Fig.~\ref{LO}, to provide the required 
phase~\cite{qiu}.
\bef
\psfig{file=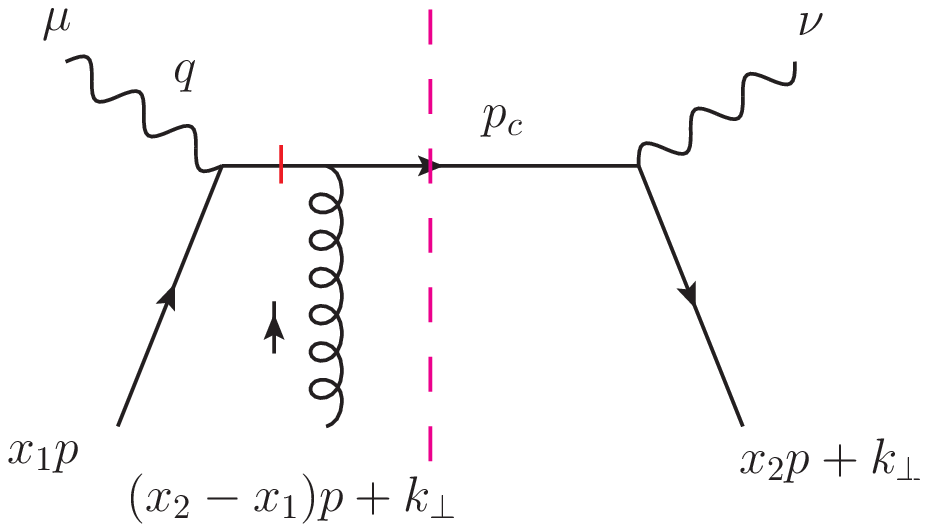, width=2.2in}
\hskip 0.3in
\psfig{file=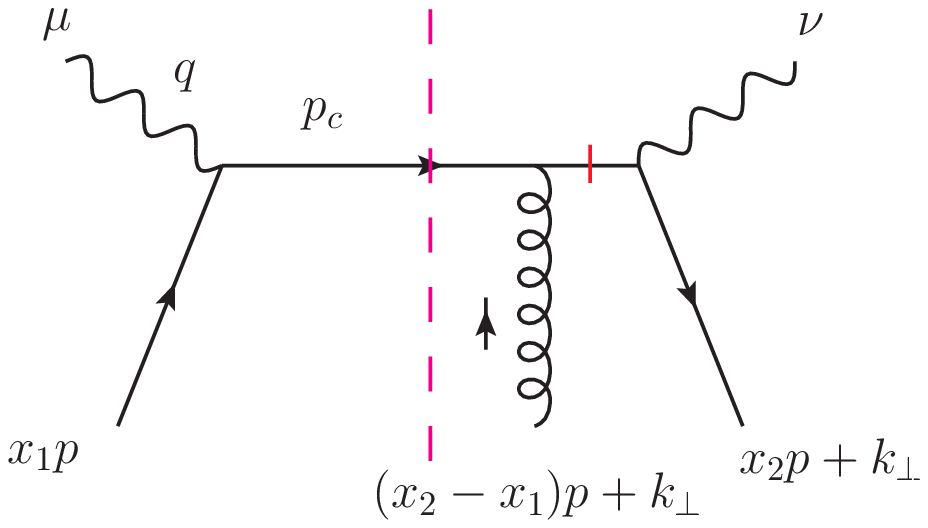, width=2.2in}
\caption{Leading order Feynman diagrams. Left: gluon to the left of the cut. Right: gluon to the right of the cut.}
\label{LO}
\eef
We  work in the covariant gauge. For the gluon on the left of the $t=\infty$  cut, shown in Fig.~\ref{LO}(left), 
we have the $\pht$-weighted cross section as
\ben
\left.\frac{d\langle \pht \Delta\sigma(S_\perp)\rangle}{dx_B dy dz_h}\right|_{\rm Fig.~\ref{LO}(left)} &\propto&
\int d^2\pht \epsilon^{\alpha\beta} S_{\perp}^\alpha \pht^{\beta} 
\int dz D_{q\to h}(z) \int dx_1 dx_2 d^2k_\perp T_{q,A}(x_1, x_2, k_\perp)
\nnu
&&
\times
\left[-g^{\mu\nu}H_{\mu\nu}(x_1, x_2, k_\perp)\right] \delta^2\left(\pht- z k_\perp \right).
\een
Here (and throughout the paper), for simplicity we only consider the so-called metric 
contribution~\cite{Altarelli:1979ub,Graudenz:1994dq,Daleo:2004pn}. This means that we contract our 
hadronic tensor $H_{\mu\nu}$ with $-g^{\mu\nu}$. The twist-3 correlation function $T_{q,A}(x_1, x_2, k_\perp)$ is defined as
\ben
T_{q,A}(x_1, x_2, k_\perp) &=& \int \frac{dy_1^-}{2\pi} \frac{dy_2^-}{2\pi}\frac{d^2y_\perp}{(2\pi)^2}
e^{ix_1P^+ y_1^-} e^{i(x_2-x_1)P^+ y_2^-} e^{ik_\perp\cdot y_\perp} 
\frac{1}{2} \langle PS| \bar{\psi}_q(0) \gamma^+ A^+(y_2^-, y_\perp) \psi_q(y_1^-)|PS\rangle.
\een
One can take advantage of $\delta^2\left(\pht- z k_\perp \right)$ to integrate out $d^2\pht$, 
thus $\pht^\beta = z k_\perp^\beta$. We then use $k_\perp^\beta$ to convert $A^+$ to the $F^{+\beta}$ 
field strength through integration by parts~\cite{Kang:2008us}. At the same time, one realizes that 
the Feynman diagram with the gluon to the right of the cut (Fig.~\ref{LO}(right)) gives no 
contribution to the $\pht$-weighted Sivers asymmetry. This is because for this diagram the 
associated $\delta$-function becomes $\delta^2(\pht)$, i.e. $\pht=0$, and the $\pht$-weighted asymmetry 
vanishes. This conclusion also holds true in the virtual diagram calculation. 

The required phase to generate a Sivers asymmetry comes from a pole in the propagator, 
which is represented by a short-bar in Fig.~\ref{LO},
\ben
\frac{1}{\left(p_c-(x_2-x_1)P\right)^2+i\epsilon} = \frac{1}{2P\cdot p_c} \frac{1}{x_1-x_2+i\epsilon}
\to \frac{1}{2P\cdot p_c} (-i\pi)\delta(x_1-x_2).
\een
With this phase, we  have
\ben
g_s \int d^2k_\perp i\epsilon^{\alpha\beta} S_\perp^\alpha k_\perp^\beta T_{q,A}(x_1, x_2, k_\perp) = 
\frac{1}{2\pi} T_{q, F}(x_1, x_2),
\een
where $T_{q, F}(x_1, x_2)$ is the well-known Qiu-Sterman function, with the following definition:
\ben
T_{q,F}(x_1, x_2) &=& \int \frac{dy_1^- dy_2^-}{4\pi}
e^{ix_1P^+ y_1^- + i(x_2-x_1)P^+ y_2^-} 
\frac{1}{2} \langle PS| \bar{\psi}_q(0) \gamma^+ 
\epsilon^{\alpha\beta}S_{\perp\alpha}F^+_{~~\beta}(y_2^-) \psi_q(y_1^-)|PS\rangle.
\een
Finally, the $\pht$-weighted Sivers asymmetry at LO has the following form~\cite{Boer:1997nt}
\ben
\frac{d\langle \pht \Delta\sigma(S_\perp)\rangle}{dx_B dy dz_h} = -\frac{z_h\sigma_0}{2} \sum_q e_q^2
\int \frac{dx}{x} \frac{dz}{z} T_{q,F}(x,x) D_{q\to h}(z) \delta(1-\hat x)\delta(1-\hat z) , 
\label{lo-res}
\een
where we recall that $\hat x= x_B/x$ and $\hat z=z_h/z$. Since we will use dimensional 
regularization for our NLO calculation in the next section, we also need the LO result in 
$n=4-2\epsilon$ dimension. We find that the only change is the appearance of $1-\epsilon$ in the 
overall normalization of $\sigma_0$, i.e. in $n=4-2\epsilon$ dimension 
we have $\sigma_0$ in Eq.~\eqref{lo-res} defined as
\ben
\sigma_0 = \frac{2\pi\alpha_{\rm em}^2}{Q^2} \frac{1+(1-y)^2}{y} (1-\epsilon).
\een

\section{Transverse Momentum-Weighted Sivers asymmetry at NLO}
In this section we present the NLO pQCD corrections to the transverse spin-dependent differential cross section. 
We first give the result for virtual corrections, and then study the real corrections. We then combine the real 
and virtual corrections to obtain the final expression. We  show that all the soft divergences cancel out between 
real and virtual diagrams. The remaining collinear divergence can be absorbed by the redefinition of the 
unpolarized fragmentation function, and the twist-3 Qiu-Sterman function. This provides an alternative way 
to derive the evolution equation for the Qiu-Sterman function.

\subsection{Virtual corrections}
\bef
\psfig{file=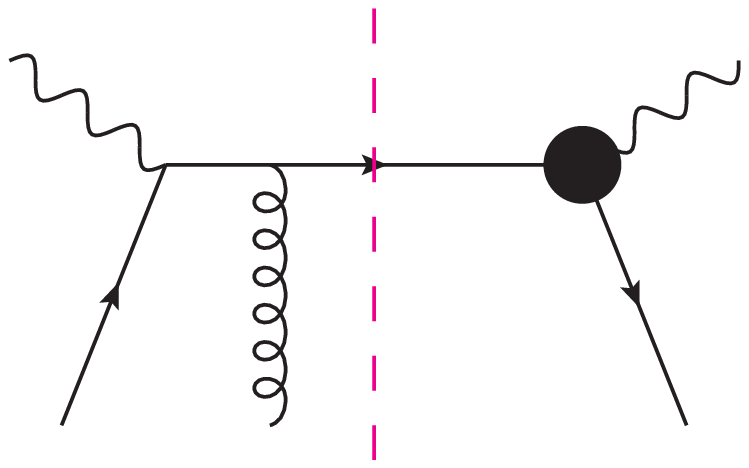, width=1.9in}
\hskip 0.3in
\psfig{file=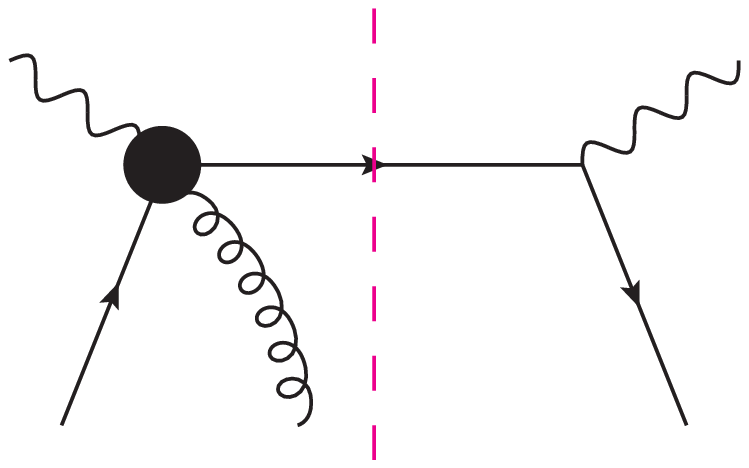, width=1.9in}
\caption{The generic Feynman diagrams for the virtual corrections to the $\pht$-weighted cross section.}
\label{virtual-generic}
\eef
We first study the virtual corrections. The relevant generic Feynman diagrams are 
shown in Fig.~\ref{virtual-generic}. Here, we only include the diagrams which have the 
gluon attached to the left of the cut. This is because the diagrams with the gluon to the 
right of the cut, just like in the LO calculation, give no contribution to the $\pht$-weighted 
Sivers asymmetry because of  the same $\delta$-function $\delta^2(\pht)$. The blob in 
Fig.~\ref{virtual-generic}(left) is given by Fig.~\ref{virtual-left-1}. These diagrams are pretty 
easy to compute since they are the same as the usual virtual corrections in the unpolarized cross 
section. The result is given by~\cite{Altarelli:1979ub}
\ben
\left.\frac{d\langle \pht \Delta\sigma(S_\perp)\rangle}{dx_B dy dz_h} 
\right|_{\rm Fig.~\ref{virtual-generic}(left)}&=& -\frac{z_h\sigma_0}{2} \frac{\alpha_s}{4\pi} \sum_q e_q^2
\int \frac{dx}{x} \frac{dz}{z} T_{q,F}(x,x) D_{q\to h}(z) \delta(1-\hat x)\delta(1-\hat z)
\nnu
&&
\times C_F \left(\frac{4\pi\mu^2}{Q^2}\right)^\epsilon \frac{1}{\Gamma(1-\epsilon)}
\left[-\frac{2}{\epsilon^2}-\frac{3}{\epsilon} - 8 \right]
\een
\bef
\psfig{file=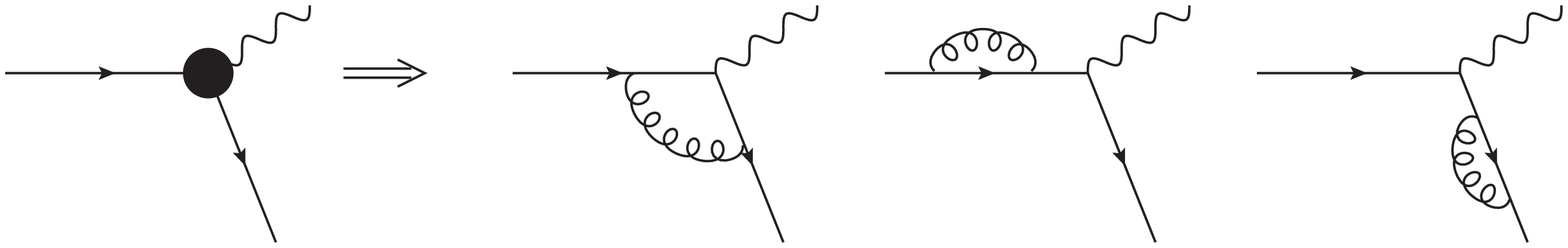, width=5.3in}
\caption{One-loop virtual corrections for the $\pht$-weighted cross section: 
shown here are the corrections to the quark-photon-quark vertex, 
corresponding to the blob in  Fig.~\ref{virtual-generic}(left).}
\label{virtual-left-1}
\eef

On the other hand, the blob in Fig.~\ref{virtual-generic}(right) is much more complicated and the explicit 
diagrams are given in Fig.~\ref{virtual-left-2}.
\bef
\psfig{file=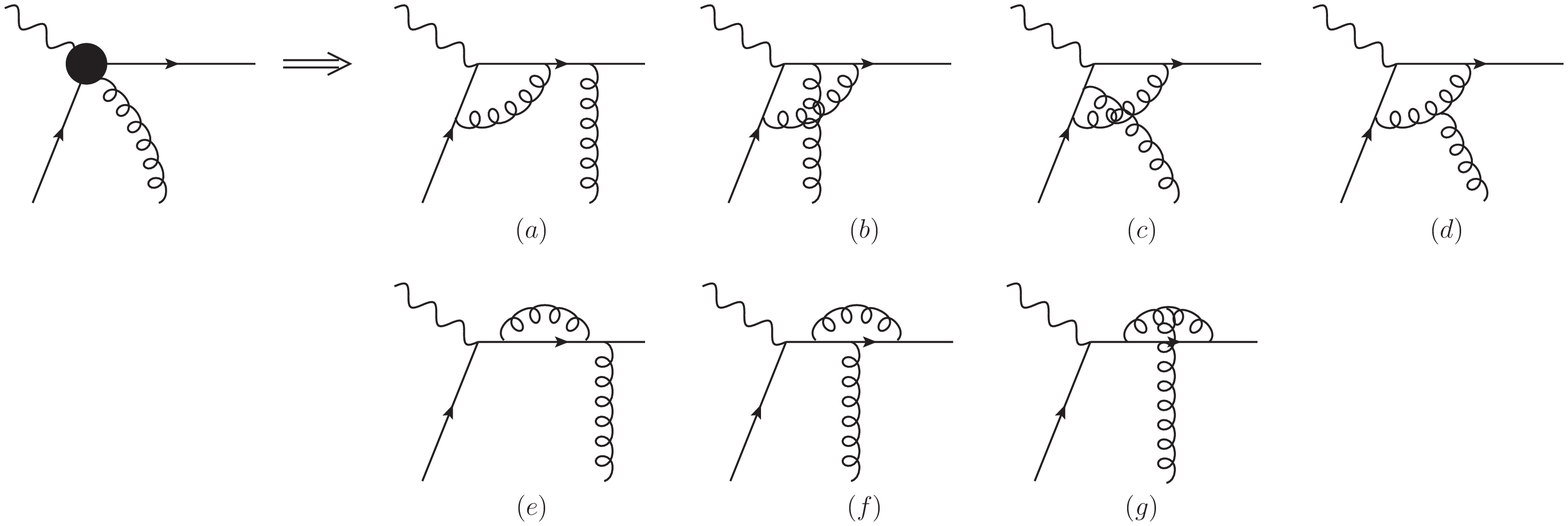, width=6.2in}
\caption{One-loop virtual corrections for the $\pht$-weighted cross section: shown here are the corrections to 
the quark-photon-quark vertex with gluon attachment, corresponding to the blob in 
 Fig.~\ref{virtual-generic}(right).}
\label{virtual-left-2}
\eef
The calculation is  lengthy  and contains significant amount of tensor reduction and integration. 
The diagrams contain three types of color factors: (a) and (e) have color factors $C_F$; (b), (c) and (f) have 
color factors $-1/2N_c =C_F- N_c/2$; (d) and (g) have color factors $N_c/2$. We find that the 
terms associated with  $N_c/2$ cancel out and only the color structure proportional to $C_F$ remains. 
The final result is 
\ben
\left.\frac{d\langle \pht \Delta\sigma(S_\perp)\rangle}{dx_B dy dz_h} 
\right|_{\rm Fig.~\ref{virtual-generic}(right)} &=& -\frac{z_h\sigma_0}{2} \frac{\alpha_s}{4\pi}  \sum_q e_q^2
\int \frac{dx}{x} \frac{dz}{z} T_{q,F}(x,x) D_{q\to h}(z) \delta(1-\hat x)\delta(1-\hat z)
\nnu
&&
\times C_F \left(\frac{4\pi\mu^2}{Q^2}\right)^\epsilon \frac{1}{\Gamma(1-\epsilon)}
\left[-\frac{2}{\epsilon^2}-\frac{3}{\epsilon} - 8 \right]
\een
Thus, the virtual correction is given by the sum of both diagrams in Fig.~\ref{virtual-generic}:
\ben
\left.\frac{d\langle \pht \Delta\sigma(S_\perp)\rangle}{dx_B dy dz_h} \right|_{\rm virtual} &=& 
-\frac{z_h\sigma_0}{2} \frac{\alpha_s}{2\pi} \sum_q e_q^2
\int \frac{dx}{x} \frac{dz}{z} T_{q,F}(x,x) D_{q\to h}(z) \delta(1-\hat x)\delta(1-\hat z)
\nnu
&&
\times  C_F \left(\frac{4\pi\mu^2}{Q^2}\right)^\epsilon \frac{1}{\Gamma(1-\epsilon)}
\left[-\frac{2}{\epsilon^2}-\frac{3}{\epsilon} - 8 \right]
\label{virtual-result}
\een

\subsection{Real corrections}
Let us  now study the real corrections. In this case, we need to perform the usual $k_\perp$-expansion 
(also referred to as collinear expansion). The techniques for $k_\perp$-expansion are well established in the 
literature, see e.g.  Refs.~\cite{qiu,koike,Koike:2011mb,Liang:2012rb}. The  $\pht$-weighted cross section 
can be written as follows:
\ben
\left.\frac{d\langle \pht \Delta\sigma(S_\perp)\rangle}{dx_B dy dz_h} \right|_{\rm Real} 
&\propto&
\int d^2\pht \epsilon^{\alpha\beta} S_{\perp}^\alpha \pht^{\beta} 
\int dz D_{q\to h}(z) \int dx_1 dx_2 d^2k_\perp T_{q,A}(x_1, x_2, k_\perp)
\nnu
&&
\times
k_\perp^\rho \frac{\partial}{\partial k_\perp^\rho}\left[-g^{\mu\nu}H_{\mu\nu}(x_1, x_2, k_\perp)\right]
\label{real}
\een
Again, we need a phase to generate the Sivers asymmetry, which also comes from the pole in the propagators. 
For the SIDIS process we can have both soft-pole and hard-pole contributions. Soft-pole contributions 
come from the Feynman diagrams shown in Fig.~\ref{soft-pole-figs}, with the soft-pole marked by a short bar 
in the diagram. For  gluon to the left of the cut, it arises from
\ben
\frac{1}{\left(p_c - (x_2-x_1)P-k_\perp\right)^2+i\epsilon} = 
\frac{1}{2P\cdot p_c}\frac{1}{x_1-x_2-v_1\cdot k_\perp+i\epsilon} 
\to -\frac{x}{\hat u} (-i\pi)\delta(x_1-x_2+v_1\cdot k_\perp) , 
\een
where the Mandelstam variables are defined as
\ben
\hat s = (xP+q)^2,
\qquad
\hat t=(p_c-q)^2,
\qquad
\hat u=(xP-p_c)^2.
\een
Here, $p_c$ is the momentum of the final-state quark which fragments into the observed hadron 
and $v_1^\mu=2xp_c^\mu /\hat u$. When $k_\perp\to 0$, $x_1=x_2$, i.e. the attached gluon momentum becomes zero. 
This clarifies the name ``soft-pole'' contribution.
\bef
\psfig{file=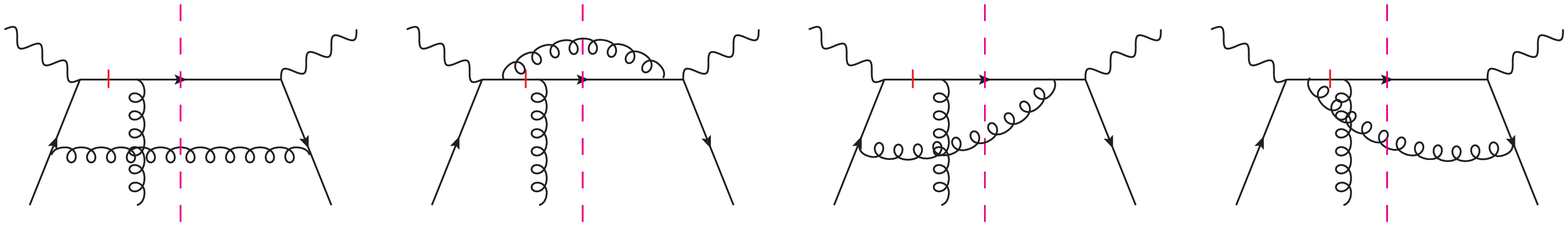, width=6.2in}
\caption{Feynman diagrams for soft-pole contributions to the $\pht$-weighted transverse spin-dependent cross section. 
The short bars indicate the propagators where the soft pole arises. The ``mirror'' diagrams 
(those with gluon attached to the right of the cut) are not shown, but are included in the calculations.}
\label{soft-pole-figs}
\eef

On the other hand, the hard-pole contributions come from the Feynman diagrams shown in 
Fig.~\ref{hard-pole-figs}, with the hard-pole marked by a short-bar in the diagram. 
For the gluon to the left of the cut, the phase arises as follows:
\ben
\frac{1}{(x_1P+q)^2+i\epsilon} \to \frac{1}{2P\cdot q}\frac{1}{x_1-x_B+i\epsilon} 
\to \frac{x}{\hat s+Q^2} (-i\pi)\delta(x_1-x_B),
\een
which sets $x_1=x_B=xQ^2/(\hat s+Q^2)$. On the other hand, the on-shell condition for the unobserved gluon 
leads to another $\delta$-function, $\delta(x_2-x-v_2\cdot k_\perp)$ with $v_2^\mu=-2xp_c^\mu/\hat t$. 
Thus when $k_\perp\to 0$, the attached gluon momentum $\sim x_2-x_1 = x - x_B = x\hat s/(\hat s+Q^2)$, 
which is finite. This clarifies the name ``hard-pole'' contribution.
\bef
\psfig{file=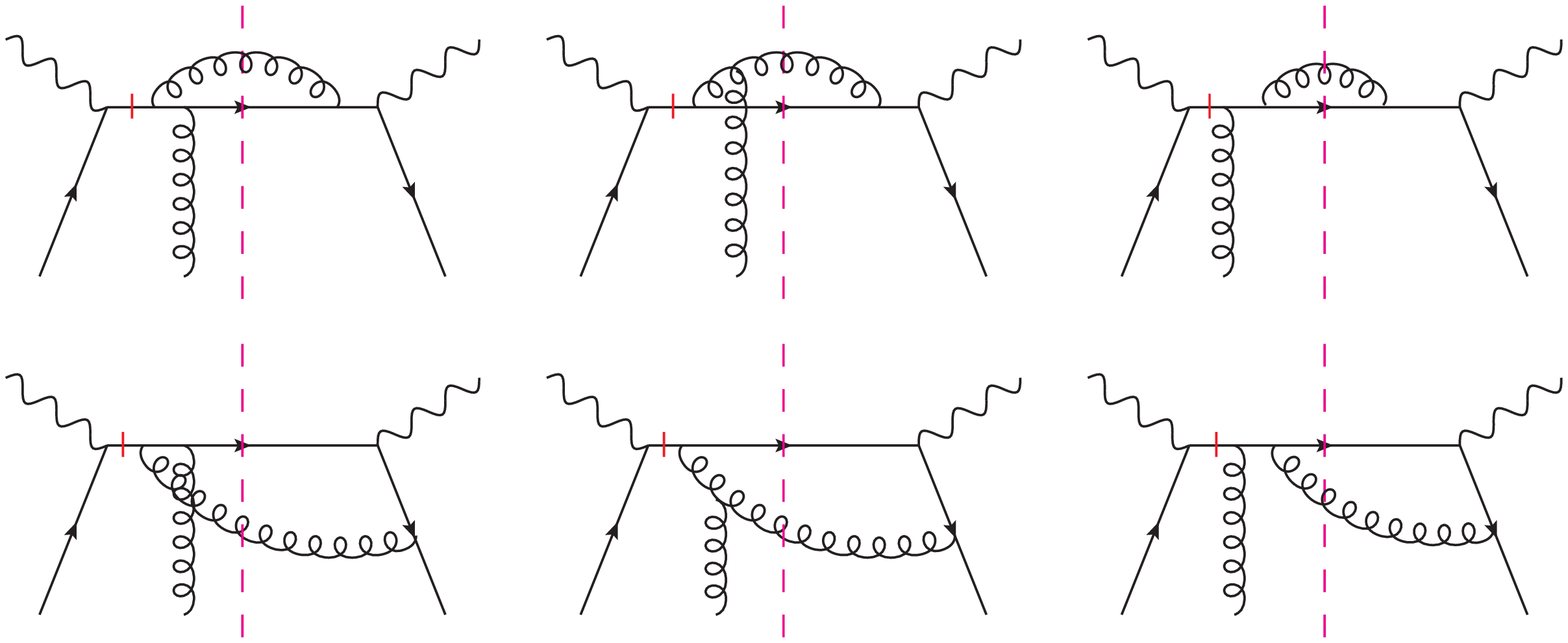, width=4.8in}
\caption{Feynman diagrams for hard-pole contributions to the $\pht$-weighted transverse spin-dependent 
cross section. The short bars indicate the propagators where the hard pole arises. The ``mirror'' 
diagrams (those with gluon attached to the right of the cut) are not shown, but are included 
in the calculations.}
\label{hard-pole-figs}
\eef

The collinear expansion for both soft-pole and hard-pole contributions are straightforward. 
After such an expansion, we will have a linear $P_{h\perp}^\rho$-dependence:
\ben
\frac{\partial}{\partial k_\perp^\rho}\left[-g^{\mu\nu}H_{\mu\nu}(x_1, x_2, k_\perp)\right] 
\propto p_{c\perp}^\rho = \frac{\pht^\rho}{z}.
\een
This linear $\pht^\rho$ will combine with $\pht^\beta$ in Eq.~\eqref{real} to become $\pht^2$,
\ben
\pht^\beta \pht^\rho &\to& \frac{1}{2(1-\epsilon)} \pht^2 g_\perp^{\beta\rho},
\een
where $g_\perp^{\beta\rho}$ is the metric tensor in the transverse components. We can further express 
$\pht^2$ in terms of Mandelstam variables,
\ben
\pht^2 &=& z^2 \frac{\hat s\hat t \hat u}{(\hat s+Q^2)^2}.
\een
Finally our result for the real corrections is given by
\ben
\left.\frac{d\langle \pht \Delta\sigma(S_\perp)\rangle}{dx_B dy dz_h} \right|_{\rm Real} 
&=&
-\frac{1}{2}\sigma_0  \sum_q e_q^2 \int \frac{dx}{x} 
dz D_{q\to h}(z) \frac{\alpha_s}{2\pi} \left(\frac{4\pi\mu^2}{Q^2}\right)^\epsilon
\frac{1}{\Gamma(1-\epsilon)} \frac{1}{1-\epsilon}
\hz^{-\epsilon} (1-\hz)^{-\epsilon}\hx^\epsilon (1-\hx)^{-\epsilon}
\nnu
&&
\times
\left[ x\frac{d}{dx} T_{q,F}(x,x) D_q^s + T_{q,F}(x,x) N_q^s + T_{q,F}(x,x_B) N_q^h\right] , 
\label{real-result}
\een
where we have used the expression for the two-body phase space integral in $n=4-2\epsilon$ dimension
\ben
dPS^{(2)} &=& \frac{d^{n-1}p_c}{(2\pi)^{n-1} 2p_c^0}  \frac{d^{n-1}p_d}{(2\pi)^{n-1} 2p_d^0} 
(2\pi)^n\delta^n(xP+q-p_c-p_d),
\nnu
&=& \frac{dz_h}{z} \frac{1}{8\pi} \left(\frac{4\pi}{Q^2}\right)^\epsilon \frac{1}{\Gamma(1-\epsilon)} 
\hz^{-\epsilon} (1-\hz)^{-\epsilon}\hx^\epsilon (1-\hx)^{-\epsilon}.
\een
The hard-part coefficient functions in Eq.~\eqref{real-result} are
\ben
D_q^s &=& \frac{1}{2N_c} \frac{\hat s}{\hat s+Q^2} 
\left[ (1-\epsilon) \left(-\frac{\hat s}{\hat t} - \frac{\hat t}{\hat s} \right)
+\frac{2Q^2\hat u}{\hat s\hat t} + 2\epsilon
\right],
\\
N_q^s &=& \frac{1}{2N_c} \frac{1}{\hat s \hat t (\hat s+Q^2)}
\left[ (\hat s+Q^2)^3+Q^2 (\hat u^2-\hat s^2)+\hat s(\hat s+\hat u)^2
-\epsilon(Q^2+\hat u)
\left((Q^2+\hat u)(Q^2+\hat s)+2Q^2\hat s\right)\right],
\\
N_q^h &=& \left[\frac{\hat u}{\hat t+\hat u} C_F + \frac{1}{2N_c}\right]
\frac{1}{\hat s \hat t (\hat s+Q^2)}
\left[(\hat t+\hat u)^3 - Q^2\hat u^2+\epsilon \hat t (Q^2+\hat u)(\hat t+\hat u)\right].
\een

\subsection{Combining the  real and virtual corrections}
To combine real and virtual corrections and demonstrate how the divergences cancel out, we need 
to perform the $\epsilon$-expansion for the real corrections in Eq.~\eqref{real-result}. To proceed, 
we realize that
\ben
\hat s = \frac{1-\hx}{\hx} Q^2, \qquad
\hat t = - \frac{1-\hz}{\hx} Q^2, \qquad
\hat u = -\frac{\hz}{\hx} Q^2. 
\een
Let us  define the following common factor: 
\ben
I=\frac{1}{1-\epsilon} \hz^{-\epsilon} (1-\hz)^{-\epsilon} \hx^\epsilon (1-\hx)^{-\epsilon} .
\een
We carry out the $\epsilon$-expansion for the $I \times (D_q^s, N_q^s, N_q^h)$ terms. Using the 
following formulas~\cite{Altarelli:1979ub}:
\ben
\hz^{-\epsilon} (1-\hz)^{-\epsilon-1} &=& -\frac{1}{\epsilon} \delta(1-\hz)+\frac{1}{(1-\hz)_+} 
- \epsilon \left(\frac{\ln(1-\hz}{1-\hz}\right)_+ - \epsilon \frac{\ln\hz}{1-\hz},
\\
\hx^{\epsilon} (1-\hx)^{-\epsilon-1} &=& -\frac{1}{\epsilon} \delta(1-\hx)+\frac{1}{(1-\hx)_+} 
- \epsilon \left(\frac{\ln(1-\hx}{1-\hx}\right)_+ + \epsilon \frac{\ln\hx}{1-\hx},
\\
\hz^{-\epsilon} (1-\hz)^{-\epsilon} &=& 1-\epsilon \ln\hz - \epsilon\ln(1-\hz),
\\
\hx^{\epsilon} (1-\hx)^{-\epsilon} &=& 1+\epsilon \ln\hx - \epsilon \ln(1-\hx),
\een
we present the expanded results for these hard-part coefficient functions. For the so-called derivative term $D_q^s$  we have
\ben
I\times D_q^s = \frac{1}{2N_c} \left[-\frac{1}{\epsilon} (1+\hx^2)\delta(1-\hz)
+(1-\hz)+\frac{(1-\hx)^2+2\hx\hz}{(1-\hz)_+} - (1+\hx^2) \ln\frac{\hx}{1-\hx}\delta(1-\hz) - 2\hx\delta(1-\hz) 
\right].
\een
For the first divergent piece we  further perform integration by parts to convert
 $x\frac{d}{dx} T_{q,F}(x,x)$ to the function $T_{q,F}(x,x)$ itself:
\ben
\frac{1}{2N_c}\int_{x_B}^1 \frac{dx}{x}\,  x\frac{d}{dx}T_{q,F}(x,x)(1+\hx^2) =\frac{1}{2N_c} \int_{x_B}^1 \frac{dx}{x} T_{q,F}(x,x)
\left[2\hx^2 - 2\delta(1-\hx)\right].
\label{error}
\een
Thus, the term associated with $D_q^s$ leads to the following expression:
\ben
&&\int\frac{dx}{x} \left(-\frac{1}{\epsilon}\right) \delta(1-\hz) T_{q,F}(x,x) \frac{1}{2N_c}\left[2\hx^2 - 2\delta(1-\hx)\right]
\nnu
&&
+x\frac{dx}{x}T_{q,F}(x,x) 
\frac{1}{2N_c} \left[(1-\hz)+\frac{(1-\hx)^2+2\hx\hz}{(1-\hz)_+} - (1+\hx^2) \ln\frac{\hx}{1-\hx}\delta(1-\hz) - 2\hx\delta(1-\hz) 
\right].
\label{dqs}
\een
Likewise, we have 
\ben
I\times N_q^s &=& \frac{1}{2N_c}\Bigg\{ - \frac{2}{\epsilon^2} \delta(1-\hx)\delta(1-\hz)  - \frac{2}{\epsilon}  \delta(1-\hx)\delta(1-\hz)
+\frac{1}{\epsilon} \frac{1+\hz^2}{(1-\hz)_+} \delta(1-\hx)
\nnu
&&
-\frac{1}{\epsilon} \frac{2\hx^3-3\hx^2-1}{(1-\hx)_+} \delta(1-\hz) - 2 \delta(1-\hx)\delta(1-\hz)
\nnu
&&
+\delta(1-\hz) \left[-(1-\hx)(1+2\hx) \ln\frac{\hx}{1-\hx}-2\left(\frac{\ln(1-\hx)}{1-\hx}\right)_+
+\frac{2}{(1-\hx)_+}  - 2(1-\hx) + 2\frac{\ln \hx}{1-\hx}\right]
\nnu
&&
+\delta(1-\hx) \left[ (1+\hz) \ln \hz(1-\hz) - 2\left(\frac{\ln(1-\hz)}{1-\hz}\right)_+
+\frac{2\hz}{(1-\hz)_+}
-2\frac{\ln \hz}{1-\hz}
\right]
\nnu
&&
+\frac{2\hx^3 - 3\hx^2-1}{(1-\hx)_+ (1-\hz)_+} + \frac{1+\hz}{(1-\hx)_+} - 2(1-\hx)\Bigg\} ,
\label{nqs}
\\
I\times N_q^h &=& \left[\hz C_F+\frac{1}{2N_c}\right]
\Bigg\{\frac{2}{\epsilon^2} \delta(1-\hx)\delta(1-\hz)  + \frac{2}{\epsilon}  \delta(1-\hx)\delta(1-\hz)
- \frac{1}{\epsilon} \frac{1+\hz^2}{(1-\hz)_+} \delta(1-\hx)
\nnu
&&
-\frac{1}{\epsilon}\frac{1+\hx}{(1-\hx)_+} \delta(1-\hz) +  2 \delta(1-\hx)\delta(1-\hz)
+\frac{1+\hx\hz^2}{(1-\hx)_+(1-\hz)_+}
\nnu
&&
+\delta(1-\hz) \left[\ln\frac{\hx}{1-\hx} + 2\left(\frac{\ln(1-\hx)}{1-\hx}\right)_+
-2\frac{\ln\hx}{1-\hx} - \frac{1+\hx}{(1-\hx)_+}
\right]
\nnu
&&
+\delta(1-\hx) \left[-(1+\hz) \ln\hz(1-\hz) + 2 \left(\frac{\ln(1-\hz)}{1-\hz}\right)_+
+2\frac{\ln\hz}{1-\hz} - \frac{2\hz}{(1-\hz)_+}
\right]\Bigg\} .
\label{nqh}
\een

Now let us combine the results for the real corrections in Eqs.~\eqref{dqs}, \eqref{nqs}, and \eqref{nqh} with the 
virtual correction in Eq.~\eqref{virtual-result}. First, for the double-pole $1/\epsilon^2$ term, which represents a 
soft-collinear divergence, we find that they cancel out between real and virtual diagrams. Specifically, the 
term $\sim 1/2N_c$ in the soft-pole contribution $N_q^s$ cancel the corresponding term in the hard-pole contribution 
$N_q^h$, which leaves the remaining $1/\epsilon^2$ term in $N_q^h$ with a color factor $C_F$. This remaining term 
exactly cancels the $1/\epsilon^2$ term in the virtual diagrams.

Now we turn our attention to the $1/\epsilon$ term. By adding the corresponding terms in real 
and virtual diagrams, we end up with the following expression:
\ben
\hz \left(-\frac{1}{\epsilon}\right)
\left[
\delta(1-\hx) T_{q,F}(x,x) P_{qq}(\hz)
+\delta(1-\hz) P_{qg\to qg}\otimes T_{q,F}(x, x\hx)
\right] , 
\een
where $P_{qq}(\hz)$ is the usual quark-to-quark splitting kernel, as in Eq.~\eqref{Pqq}. Thus, the first term, the part containing $\hz$, is just the collinear 
QCD correction to the bare leading order fragmentation function $D_{q\to h}^{(0)}(z_h)$ (after including the 
pre-factor $(4\pi\mu^2/Q^2)^\epsilon/\Gamma[1-\epsilon]$):
\ben
D_{q\to h}(z_h) = D_{q\to h}^{(0)}(z_h)
+\frac{\alpha_s}{2\pi} \int_{z_h}^1 \frac{dz}{z} \left(-\frac{1}{\hat{\epsilon}}\right) D_{q\to h}(z) P_{qq}(\hz) ,
\een
where we have adopted $\overline{\rm MS}$-scheme, and 
\ben
\frac{1}{\hat\epsilon} = \frac{1}{\epsilon} - \gamma_E + \ln4\pi.
\een
On the other hand, the second term $-\frac{1}{\epsilon}P_{qg\to qg}\otimes T_{q,F}(x, x\hx)$ is given by
\ben
\left(-\frac{1}{\epsilon}\right)P_{qg\to qg}\otimes T_{q,F}(x, x\hx) &=& \left(-\frac{1}{\epsilon}\right)
\Bigg\{T_{q,F}(x,x)
C_F \left[\frac{1+\hx^2}{(1-\hx)_+}+\frac{3}{2}\delta(1-\hx)\right] -N_c\delta(1-\hx) T_{q,F}(x,x)
\nnu
&&
+\frac{N_c}{2}\left[\frac{1+\hx}{(1-\hx)_+}T_{q,F}(x, x\hat{x}) - \frac{1+\hx^2}{(1-\hx)_+} T_{q,F}(x,x) \right]\Bigg\},
\een
from which we immediately obtain the collinear QCD correction to the leading order bare Qiu-Sterman function $T_{q,F}^{(0)}(x_B,x_B)$ as follows:
\ben
T_{q,F}(x_B,x_B) &=& T_{q,F}^{(0)}(x_B,x_B) + \frac{\alpha_s}{2\pi} \int_{x_B}^1 \frac{dx}{x} \left(-\frac{1}{\hat \epsilon}\right) 
\Bigg\{
T_{q,F}(x,x)
C_F \left[\frac{1+\hx^2}{(1-\hx)_+}+\frac{3}{2}\delta(1-\hx)\right] -N_c\delta(1-\hx) T_{q,F}(x,x)
\nnu
&&
+\frac{N_c}{2}\left[\frac{1+\hx}{(1-\hx)_+}T_{q,F}(x, x\hat{x}) - \frac{1+\hx^2}{(1-\hx)_+} T_{q,F}(x,x) \right]
\Bigg\}.
\een
From this equation, we can obtain the evolution equation for the twist-3 Qiu-Sterman function as follows
\ben
\frac{\partial}{\partial \ln\mu^2} T_{q,F}(x_B,x_B,\mu^2)
&=&\frac{\alpha_s}{2\pi} \int_{x_B}^1 \frac{dx}{x} 
\Bigg\{
T_{q,F}(x,x,\mu^2)
C_F \left[\frac{1+\hx^2}{(1-\hx)_+}+\frac{3}{2}\delta(1-\hx)\right] -N_c\delta(1-\hx) T_{q,F}(x,x,\mu^2)
\nnu
&&
+\frac{N_c}{2}\left[\frac{1+\hx}{(1-\hx)_+}T_{q,F}(x, x\hat{x}, \mu^2) - \frac{1+\hx^2}{(1-\hx)_+} T_{q,F}(x,x,\mu^2) \right]\Bigg\}.
\een
This result agrees with earlier findings from different approaches~\cite{Braun:2009mi,Kang:2012em, Schafer:2012ra,Ma:2012xn}. 
In particular, we are able to reproduce the term $-N_c T_{q,F}(x,x,\mu^2)$, which was missing in some early
 works~\cite{Kang:2008ey,Zhou:2008mz,Vogelsang:2009pj}. This piece 
arises as follows: there is a boundary term $\propto \frac{2}{\epsilon} \frac{1}{2N_c} \delta(1-\hx)$ from $D_q^s$, as in Eqs.~\eqref{error} and \eqref{dqs}, 
and this term cancels the same term with opposite sign in $N_q^s$ in Eq.~\eqref{nqs}. On the other hand, the 
hard-pole contribution $N_q^h$ contains the following term:
\ben
\left[\hz C_F+\frac{1}{2N_c}\right] \frac{2}{\epsilon} \delta(1-\hx)\delta(1-\hz) 
=\left(C_F+\frac{1}{2N_c} \right) \frac{2}{\epsilon} \delta(1-\hx)\delta(1-\hz) 
= \left(-\frac{1}{\epsilon}\right)\left[-N_c \delta(1-\hx)\delta(1-\hz) \right],
\een
which gives exactly the $-N_c T_{q,F}(x,x,\mu^2)$ term to the evolution of the Qiu-Sterman function $T_{q,F}(x,x,\mu^2)$.

Now, let us turn to the finite NLO corrections to the hard-part coefficient function. After $\overline{\rm MS}$ subtraction 
of the collinear divergences into the fragmentation function $D_{q\to h}(z, \mu^2)$ and the twist-3 Qiu-Sterman function 
$T_{q,F}(x,x,\mu^2)$, we obtain the NLO correction for both soft-pole and hard-pole contributions to the $\pht$-weighted 
transverse spin-dependent cross section:
\ben
\frac{d\langle \pht \Delta\sigma(S_\perp)\rangle}{dx_B dy dz_h}
&=&
-\frac{z_h\sigma_0}{2} \sum_q e_q^2
\int_{x_B}^1 \frac{dx}{x} \int_{z_h}^1 \frac{dz}{z} T_{q,F}(x,x, \mu^2) D_{q\to h}(z, \mu^2) \delta(1-\hat x)\delta(1-\hat z)
\nnu
&&
-\frac{z_h\sigma_0}{2} \frac{\alpha_s}{2\pi}  \sum_q e_q^2
\int_{x_B}^1 \frac{dx}{x} \int_{z_h}^1 \frac{dz}{z}
D_{q\to h}(z, \mu^2)
\Bigg\{
\ln\left(\frac{Q^2}{\mu^2}\right) \left[\delta(1-\hx) T_{q,F}(x,x,\mu^2) P_{qq}(\hz)  
\right.
\nnu
&&
\left.
+ \delta(1-\hz) P_{qg\to qg} \otimes T_{q,F}(x, x \hx, \mu^2) \right]
\nnu
&&
+x\frac{d}{dx} T_{q,F}(x,x,\mu^2) \frac{1}{2N_c}
\left[\frac{1-\hz}{\hz}+\frac{(1-\hx)^2+2\hx\hz}{\hz(1-\hz)_+} - \delta(1-\hz)
\left((1+\hx^2)\ln\frac{\hx}{1-\hx}+2\hx\right)
\right]
\nnu
&&
+T_{q,F}(x,x,\mu^2)\delta(1-\hz)\frac{1}{2N_c} 
\left[(2\hx^2-\hx-1)\ln\frac{\hx}{1-\hx} - 2 \left(\frac{\ln(1-\hx)}{1-\hx}\right)_+
+\frac{2\hx(2-\hx)}{(1-\hx)_+}+ 2\frac{\ln\hx}{1-\hx}
\right]
\nnu
&&
+T_{q,F}(x,x,\mu^2)\delta(1-\hx)C_F
\left[-(1+\hz)\ln\hz(1-\hz)+2 \left(\frac{\ln(1-\hz)}{1-\hz}\right)_+
-\frac{2\hz}{(1-\hz)_+}+ 2\frac{\ln\hz}{1-\hz}
\right]
\nnu
&&
+T_{q,F}(x,x,\mu^2)\frac{1}{2N_c\hz}
\left[\frac{2\hx^3-3\hx^2-1}{(1-\hx)_+(1-\hz)_+}
+\frac{1+\hz}{(1-\hx)_+} - 2(1-\hx)
\right]
\nnu
&&
+T_{q,F}(x,x \hx,\mu^2)\delta(1-\hz)\frac{N_c}{2}
\left[\ln\frac{\hx}{1-\hx}+2\left(\frac{\ln(1-\hx)}{1-\hx}\right)_+
-2\frac{\ln\hx}{1-\hx} - \frac{1+\hx}{(1-\hx)_+}
\right]
\nnu
&&
+T_{q,F}(x,x \hx,\mu^2) \frac{1+\hx\hz^2}{(1-\hx)_+(1-\hz)_+}\left(C_F+\frac{1}{2N_c\hz}\right)
-T_{q,F}(x,x,\mu^2) 6 C_F \delta(1-\hx)\delta(1-\hz)
\Bigg\}
\label{spin-dep}
\een
Just like the NLO correction to the spin-averaged cross section in Eq.~\eqref{spin-avg}, the logarithms 
containing the factorization scale together with the splitting functions determine the evolution of the twist-3 Qiu-Sterman 
function and the usual unpolarized quark-to-hadron fragmentation function. Eq.~\eqref{spin-dep} is the main result of our 
paper: once combined with the NLO spin-averaged cross section in Eq.~\eqref{spin-avg}, one will be able to compute the 
$\pht$-weighted Sivers asymmetry as defined in Eq.~\eqref{sivers}.

\section{Conclusions}
We calculated the next-to-leading order perturbative QCD corrections to the transverse momentum-weighted Sivers 
asymmetry in semi-inclusive hadron production in lepton-proton deep inelastic scattering. Specifically, we
demonstrated in detail how to evaluate at NLO the $\pht$-weighted transverse spin-dependent differential 
cross section. We found that the result can be written as a convolution of a twist-3 quark-gluon correlation function 
(often referred as Qiu-Sterman function), the usual unpolarized fragmentation function  and the hard coefficient function. 
In the course of this calculation we showed that the soft divergences cancel out between real and virtual contributions, and 
that the collinear divergences can be absorbed into the NLO twist-3 Qiu-Sterman function of the transversely polarized 
proton and the unpolarized quark-to-hadron fragmentation function. Such a procedure also provides an alternative  way to 
identify the evolution equation for the twist-3 Qiu-Sterman function. We found that our evolution equation agrees 
with those derived previously from different approaches. Using our NLO results for both the spin-averaged and 
$\pht$-weighted transverse spin-dependent differential cross section, we plan  to study in the future the $\pht$-weighted 
Sivers asymmetry. We anticipate that our findings will have important phenomenological applications relevant to 
the experimental programs at  HERMES, COMPASS, and Jefferson Lab.

\section*{Acknowledgments}
We thank Yan-Qing Ma for providing us with his Mathematica package, which is used to calculate the Feynman diagrams in the paper. 
This research is supported by the US Department of Energy, Office of Science, under Contract No.~DE-AC52-06NA25396, and 
in part by the LDRD program at LANL and NSFC of China under Project No. 10825523.

\end{document}